\documentclass{llncs}
\usepackage{todonotes}
\usepackage{paralist}
\usepackage{booktabs}
\usepackage{tabularx}

\begin{document}

\title{Should I Stay or Should I Go?}
\subtitle{On Forces that Drive and Prevent MBSE Adoption in the Embedded Systems Industry}
\author{Andreas Vogelsang\inst{1} 
\and Tiago Amorim\inst{1}
\and Florian Pudlitz\inst{1}
\and Peter Gersing\inst{2}
\and Jan Philipps\inst{3}
}

\institute{
Technische Universit{\"a}t Berlin, Germany\\
\email{\{andreas.vogelsang,buarquedeamorim,florian.pudlitz\}@tu-berlin.de},\\ 
\and
GPP Communication GmbH \& Co. KG\\
Munich, Germany\\
\email{p.gersing@gppag.de}
\and
foqee GmbH\\
Munich, Germany\\
\email{philipps@foqee.de}
}

\maketitle              

\begin{abstract}
[Context] Model-based Systems Engineering (MBSE) comprises a set of models and techniques that is often suggested as solution to cope with the challenges of engineering complex systems.
Although many practitioners agree with the arguments on the potential benefits of the techniques, companies struggle with the adoption of MBSE. [Goal] In this paper, we investigate the forces that prevent or impede the adoption of MBSE in companies that develop embedded software systems. We contrast the hindering forces with issues and challenges that drive these companies towards introducing MBSE. [Method] Our results are based on 20 interviews with experts from 10 companies. Through exploratory research, we analyze the results by means of thematic coding. 
[Results] Forces that prevent MBSE adoption mainly relate to immature tooling, uncertainty about the return-on-investment, and fears on migrating existing data and processes. On the other hand, MBSE adoption also has strong drivers and participants have high expectations mainly with respect to managing complexity, adhering to new regulations, and reducing costs.  
[Conclusions] We conclude that bad experiences and frustration about MBSE adoption originate from false or too high expectations. Nevertheless, companies should not underestimate the necessary efforts for convincing employees and addressing their anxiety. 
\end{abstract}
\keywords{System engineering, model-based, process improvement, embedded systems, interview study, empirical research}

\section{Introduction}
Model-based Systems Engineering (MBSE) describes the use of models and model-based techniques to develop complex systems, which are mainly driven by software~\cite{SPESBook}. MBSE tackles the complexity of those systems through an interrelated set of models, which connects development activities and provides comprehensive analyses.
Many companies face problems with the increasing complexity of software-intensive systems, their interdisciplinary development, and the huge amount of mainly text-based specifications. MBSE offers a solution to managing these problems and companies are attracted to its benefits.

Despite the envisioned MBSE benefits, companies are struggling with implementing it within the organization. Of course, organizational change is never easy~\cite{Change}, however other methodologies, such as agile practices, have been adopted much faster. So, what are the reasons and factors that prevent or impede companies from adopting MBSE?

In this paper, we investigate the forces that prevent or impede the adoption of MBSE in companies that develop embedded systems. We contrast forces that hinder its adoption with forces that drive companies towards introducing MBSE.

Our results are based on 20 interviews with experts from 10 organizations in Germany. We analyze the results by means of thematic coding and categorize the identified forces into inertia and anxiety forces, which prevent MBSE adoption, as well as push and pull forces, which drive the companies towards MBSE adoption. We frame the results with a coding of what the interviewees considered as MBSE.
Our paper makes the following contributions:
\begin{compactitem}
  \item We present a set of hindering and fostering forces on MBSE adoption in industry. These results were extracted from interviews with 20 experts from 10 organizations located in Germany.
  \item We analyze these forces to differentiate between MBSE specific forces and forces inherent to any kind of methodological change.
\end{compactitem}

\noindent Forces that prevent MBSE adoption mainly relate to immature tooling, uncertainty about the return-on-investment, and fears on migrating existing data and processes. On the other hand, MBSE adoption also has strong drivers and participants have high expectations mainly with respect to managing complexity, adhering to new regulations, and detecting bugs earlier. 
We observed that the hindering forces are much more concrete and MBSE-specific compared with the fostering forces, which are oftentimes very generic (e.g., increase in product quality, managing complexity, supporting reuse). Oftentimes, the interviewees could not even tell why or which part of MBSE contributes to the expected benefits. \looseness=-1

From this, we conclude that bad experiences and frustration about MBSE adoption originate from false or too high expectations. Nevertheless, companies should not underestimate the necessary efforts for convincing employees and addressing their anxiety. 

\section{Background and Related Work}
\noindent\textbf{Model-based Systems Engineering (MBSE)}
is a methodology to develop systems with focus on models. Compared with traditional development, MBSE supports engineers with automation capabilities (e.g., code generation, document derivation) and enhanced analysis capabilities (e.g., behavioral analysis, performance analysis, simulation). INCOSE defines MBSE as the following~\cite{INCOSE.2007}:
\emph{``MBSE is the formalized application of modeling to support system requirements, design, analysis, verification and validation activities beginning in the conceptual design phase and continuing throughout development and later life cycle phases.''}


 
UML and SysML are standardized graphical modeling languages for MBSE with capabilities to define different types of models, processes, procedures, and operations. 
While UML is predominantly used for software development, SysML encompasses also physical aspects of a system. The languages' graphical models are intended to cover all development phases of a system.

In some application domains, MBSE is widely used and is an integral part of development~\cite{Bone.2010}. Large tool vendors, such as IBM, Oracle, Microsoft, or the Eclipse Foundation offer tooling solutions for MBSE.

\noindent\textbf{Studies on MBSE Adoption.}
Bone~and~Cloutier~\cite{Bone.2010} report on a survey conducted by the OMG, in which participants were asked about MBSE adoption within their organization. 
\emph{Culture and general resistance to change} was identified in the study as the largest inhibitor for MBSE adoption. 
The study found that SysML is being used primarily for large-scale systems.

Motamedian~\cite{Motamedian.2013} performed an applicability analysis for MBSE. Similar to the results of Bone and Cloutier, she found that MBSE is widely used in specific application areas. She reported that 50--80\% of respondents who declared the use of MBSE in real programs or projects work in defense and aircraft industries. In contrast, over all responses, only 10\% of participants claimed that they use MBSE in their organization. 
The study identified \emph{lack of related knowledge and skills} as main barrier to MBSE introduction.

Mohagheghi~et~al.~\cite{Mohagheghi2013} collected data from four large companies that use \emph{Model-Driven Engineering (MDE)} in different projects. Their study summarizes qualitative data from internal empirical studies, interviews, and a survey to investigate the state of the practice and adoption of MDE. All participants see advantages in \emph{developing domain-specific solutions and modeling at different levels of abstraction}. None of the companies mentioned \emph{shorter development time} or \emph{improved quality of code} as main motivation. In addition, the integration with other tools is problematic and mature tools for complex models are missing. 
\emph{Higher degree of automation and reuse} was considered the most important aspect to improve productivity in the long-term.
Hutchinson~et~al.~\cite{Hutchinson:2011} describe the practices of three commercial organizations as they adopted MBSE. Later, they built a taxonomy of tool-related issues affecting the adoption of MBSE~\cite{Whittle:2017}.

Kuhn~et~al.~\cite{kuhn2012exploratory} focus on contextual forces and frictions of MBSE adoption in large companies. They found that \emph{diffing in product lines}, \emph{problem-specific languages and types}, \emph{live modeling}, and \emph{traceability between artifacts} are the main drivers for adopting MBSE. Aranda~et~al.~\cite{aranda2012transition} focus more on developers and infrastructure changes. They conclude that MBSE brings developers closer, disrupts organizational structures, and achieves improvements in productivity. 


Besides these studies on MBSE adoption, several case studies exist on applying model-based techniques to complex systems in different domains (e.g., railway~\cite{Boehm14}, automotive~\cite{Vogelsang15}, maritime traffic~\cite{Vogelsang14}).


\noindent \textbf{Summary.} Related studies report on successful applications of MBSE in several cases but also mention challenges related to its adoption. MBSE techniques are widely used in some industries, however, the majority of companies do not apply MBSE. The goal of our study is to identify reasons and forces that prevent companies from adopting MBSE and contrast them with the envisioned benefits that drive the companies towards MBSE. 

\section{Study Approach}
\subsection{Research Questions}
We structure our research by two research questions that focus on hindering and fostering forces of MBSE adoption.
\begin{compactitem}
\item RQ1: What are perceived forces that prevent MBSE adoption in industry?
\begin{itemize}
	\item RQ1.1: What are habits and inertia that prevent MBSE adoption?
	\item RQ1.2: What are anxiety factors that prevent MBSE adoption?
\end{itemize}
\item RQ2: What are perceived forces that foster MBSE adoption in industry?
\begin{itemize}
	\item RQ2.1: What are perceived issues that push industry towards MBSE?
	\item RQ2.2: What MBSE benefits are perceived as most attractive?
\end{itemize}
\end{compactitem}

\subsection{Research Design}
This is an \emph{exploratory research}~\cite{ShiRan13} based on semi-structured interviews. The method provides insights into the examined topic and gives essential information to understand the phenomenon in its real context~\cite{Dresch15,Runeson08}. We developed an interview guide~\cite{bryman15} that was structured along a funnel model~\cite{Runeson08} starting with general questions about the participant's context and the understanding of MBSE and afterwards going into detail about specific topics such as employee training, MBSE integration, or experiences in the past. 

\subsection{Data Collection and Analysis}
\textbf{Study Participants.} 
The interview participants were selected from personal contacts of the authors and industrial partners that participate in a German research project\footnote{\url{https://spedit.in.tum.de/}} that has a focus on MBSE adoption in practice. 
The interviewee selection was based on two criteria: First, the interviewee should have a work experience of several years.
Second, the interviewee should work in an environment where MBSE adoption is a realistic option. In our case, we therefore restricted the group of interviewees to people working on embedded systems or in the context of embedded systems.
It was not necessary that interviewees have adopted MBSE in their context, however, 13 of the 20 interviewees stated that they already have experiences in adopting MBSE.
Table~\ref{tbl:participants} provides an overview of the participants and their context. 
The interviews were conducted by two of the authors from May to December 2016.

\begin{table}
\centering
\caption{Study Participants}
\label{tbl:participants}
\begin{tabular}{p{1cm}p{2.4cm}p{2cm}p{4.8cm}p{1.4cm}}
\toprule
\textbf{ID}&\textbf{Industry \newline Sector}&\textbf{Type of \newline Company}&\textbf{Role of \newline Participant} & \textbf{MBSE \newline Attitude}\\
\midrule
P1 & Tool vendor & OEM & Technical Sales & neutral\\
P2 & Tool vendor & Academic & Professor & neutral\\
P3 & R\&D services& SME & Manager & neutral\\
P4 & Automotive & OEM & Head of Development & positive\\
P5 & Automotive & OEM & Systems Engineer & neutral\\
P6 & Medical & SME & Head of SW Development & positive\\
P7 & Medical & SME & Head of QA & positive\\
P8 & Automotive & Supplier & Function Architect & negative\\
P9 & Automotive & OEM & SW Architect & neutral\\
P10 & Automotive & OEM & Function Architect & positive\\
P11 & Research & Academic & Professor & negative\\
P12 & Avionics & Supplier & Technical Project Manager & neutral\\
P13 & Automotive & Supplier & Developer & positive\\
P14 & Avionics & OEM & SW Developer & neutral\\
P15 & Avionics & Supplier & SW Developer & negative\\
P16 & Avionics & OEM & Team Lead & neutral\\
P17 & Electronics & OEM & Head of SW Development & neutral\\
P18 & Avionics & SME & Head of System Engineering & negative\\
P19 & Robotics & OEM & Team Lead & positive\\
P20 & Automotive & OEM & Research and Development & negative\\
\bottomrule
\end{tabular}
\end{table} 

\textbf{Interviews.} 
There were 20 fact-to-face interviews. 
Every interview took around one hour.
In consent with the interviewee, the interviewer took notes for detailed analysis. All interview notes were managed using the qualitative data analysis tool ATLAS.ti\footnote{\url{http://atlasti.com}}.

\textbf{Analysis.}
Three researchers analyzed the interviews using \emph{qualitative coding}~\cite{Neuman.2010}. Neither of them participated in the interview phase. The study was framed using the framework of \emph{Forces on MBSE Adoption} (see Section~\ref{forces}) with the following codes: \{Push, Pull, Inertia, Anxiety\}. The analysis started with all three researchers working on the same five interviews. The results were later discussed and merged in a meeting. The discussions helped to homogenize the understanding of the codes among the researchers~\cite{Weston2001} (i.e., what\slash how to look for on each force).
The remaining 15 interviews were tackled in a cross-analysis fashion. The interviews were divided equally into three groups (A, B, C) and each researcher coded the interview transcripts of two groups (i.e., AB, BC, or AC) individually the same way as before. Then, each researcher merged the results and judged existing conflicts of the group he did not work on (a researcher coding interviews of groups AB merged the results of interviews of group C). In a round with all three researchers, the unresolved conflicts were ironed out. Finally, the codes were divided into three groups \{Pull, Inertia, (Anxiety, Push)\} and each researcher worked on the quotations of codes of a group individually, performing open coding to create second level codes. We present the results in Section~\ref{results} by reporting the codes with the number of related quotations and the number of interviews in which the code appeared. The number of quotations indicates the significance of a code over all interviews and the number of interviews indicates the pervasiveness of the code within the interviews.

\textbf{Availability of Data.} 
Due to unreasonable effort necessary for anonymizing the interview transcripts, we do not disclose them. However, we disclose the interview guide and the codebook.\footnote{\url{https://doi.org/10.6084/m9.figshare.5368453}}

\section{Results} \label{results}

\subsection{Overview and Definition of MBSE}
\begin{figure}
\centering
\includegraphics[width=0.6\columnwidth]{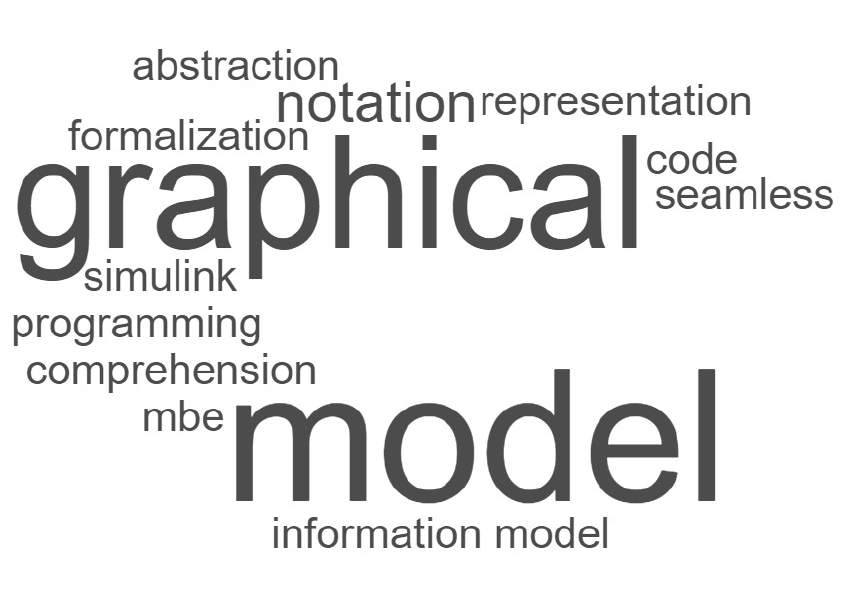}
\caption{Word cloud of MBSE descriptions}
\label{fig:def_cloud}
\end{figure}

As depicted in Table~\ref{tbl:participants}, we had a balanced set of participants with respect to MBSE attitude. For 9 out of the 20 interviews, we coded a similar number of fostering and hindering forces (i.e., neutral attitude). In 6 interviews, the fostering forces dominated (i.e., positive attitude) and in 5 interviews, the hindering forces dominated (i.e., negative attitude).
In the interviews, we did not refer to any specific MBSE approach. We did this on purpose to identify forces independent from any concrete technique or tooling. Additionally, comparing the results would have been much harder due to the large variety of MBSE approaches and flavors. Nevertheless, we asked the interviewees to define MBSE. The result can be seen in \figurename~\ref{fig:def_cloud}, where a word cloud representation of terms mentioned more than 2 times is depicted.
 
The word cloud shows the close association of MBSE with graphical models. Especially graphical descriptions of architectures and processes were mentioned several times. However, some interviewees mentioned that \emph{``graphical representation is only a part of MBSE, not everything'' (P12)} and others pointed out that MBSE should not be deformed to \emph{graphical programming}.
The only reference to a specific instance of MBSE in the word cloud is given by \emph{Simulink}.  Simulink\footnote{\url{https://de.mathworks.com/products/simulink.html}} is a widely used tool in the embedded systems domain for modeling, simulating, and analyzing dynamic systems. Interestingly, the interviewees mentioned that using Simulink is \emph{not} considered as doing MBSE (e.g., P4:\emph{``Pure implementation with Simulink is graphical programming, not MBSE.''}, P16:\emph{``Simulink is model-based engineering but not model-based \emph{systems} engineering''}). UML\slash SysML, which we expected to appear more often in the characterization of MBSE, was only mentioned rarely, however, \emph{notation} was mentioned several times. The term \emph{information model} was used a few times as important part of an MBSE approach. P7: \emph{``A core topic of MBSE is the information model that specifies and relates all development artifacts.''}
Apart from that, the interviewees frequently mentioned several well-known properties related to MBSE such as \emph{abstraction, formalization, and comprehension}.
In summary, the results show that our interviewees were not biased by a specific MBSE flavor or approach that they previously had in mind when answering our questions. However, the variety of answers also shows that the term MBSE is still far away from common understanding. 

\subsection{Forces on MBSE Adoption} \label{forces}
\begin{figure}
\centering
\includegraphics[width=0.6\columnwidth]{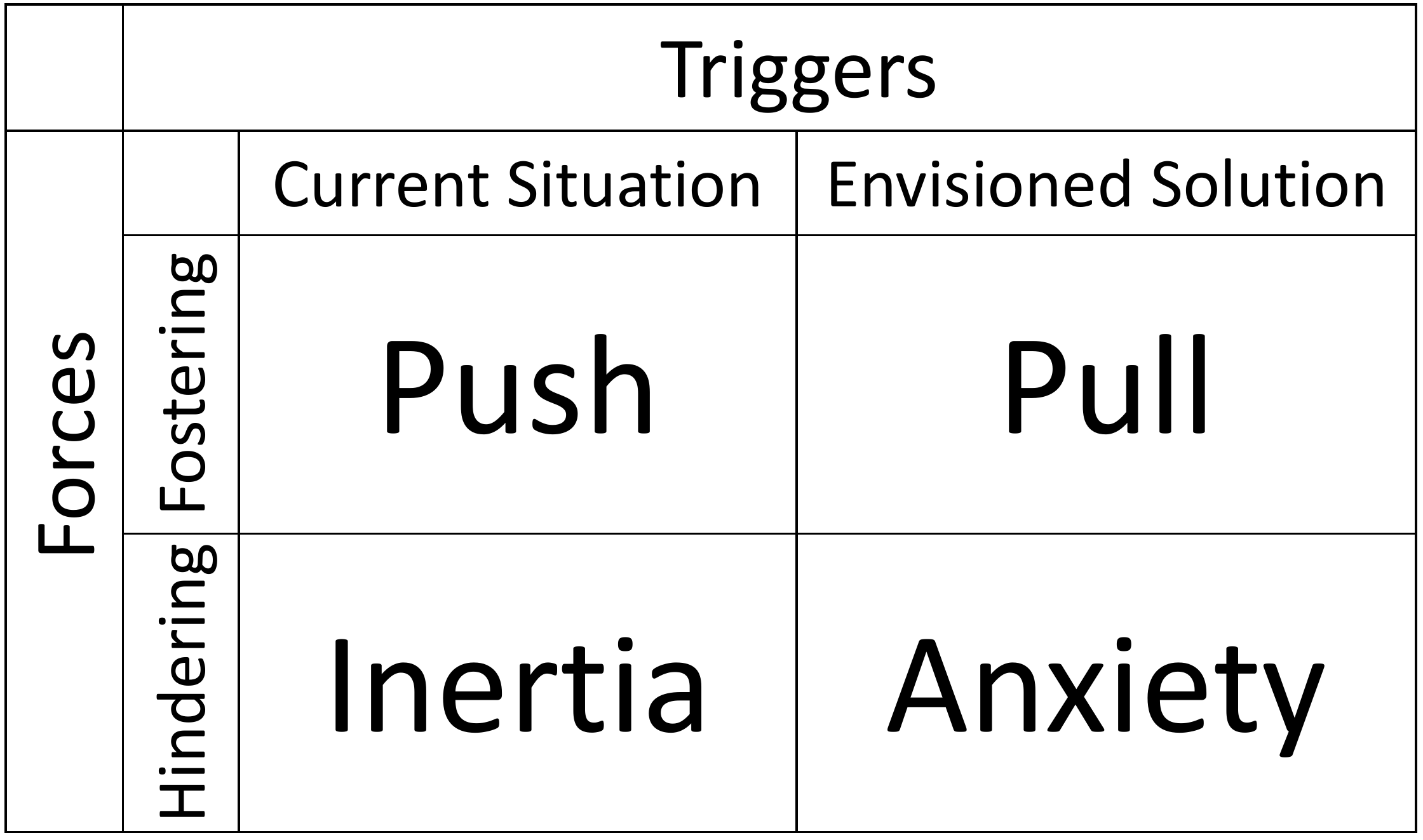}
\caption{MBSE adoption forces diagram}
\label{fig:MBSEForcesDiagram}
\end{figure}

Inspired by the categorization of Hohl~et~al.~\cite{HohlMSS16}, we defined a quadrant-wise framework for  categories of forces on MBSE adoption (see \figurename~\ref{fig:MBSEForcesDiagram}). 
The categorization aims to better understand the different aspects of the transition process from traditional to MBSE practices. We designed the framework to identify \emph{Forces} that work towards \emph{Hindering} or \emph{Fostering} the adoption of MBSE and their origin. These forces have different origins or \emph{Triggers} and are classified either into shortcomings of the \emph{Current Situation} or expected benefits of the \emph{Envisioned Solution} (MBSE in our case).
We distinguish between \emph{Push} and \emph{Pull} as forces that foster MBSE adoption. The former is triggered by issues or demands that the current situation cannot address, the latter is triggered by the ``to-be harvested" benefits of the new solution. In contrast, we define \emph{Inertia} and \emph{Anxiety} as forces that hinder MBSE adoption. The former is triggered by the feeling that the current solution is ``good enough" and habits that keep people from trying out something new. The latter is triggered by fears that MBSE introduction will not pay-off, mainly caused by uncertainties and perception flaws. 
According to Hohl~et~al.~\cite{HohlMSS16}, this classification is inspired by the Customer Forces Diagram by Maurya\footnote{\url{https://leanstack.com/science-of-how-customers-buy/}} that itself is inspired by the Forces Diagram by Moesta and Spiek from the Jobs-to-be-done framework\footnote{http://jobstobedone.org}. All four forces are present within an organization at the same time. 

\begin{figure}
\centering
\includegraphics[width=0.6\columnwidth]{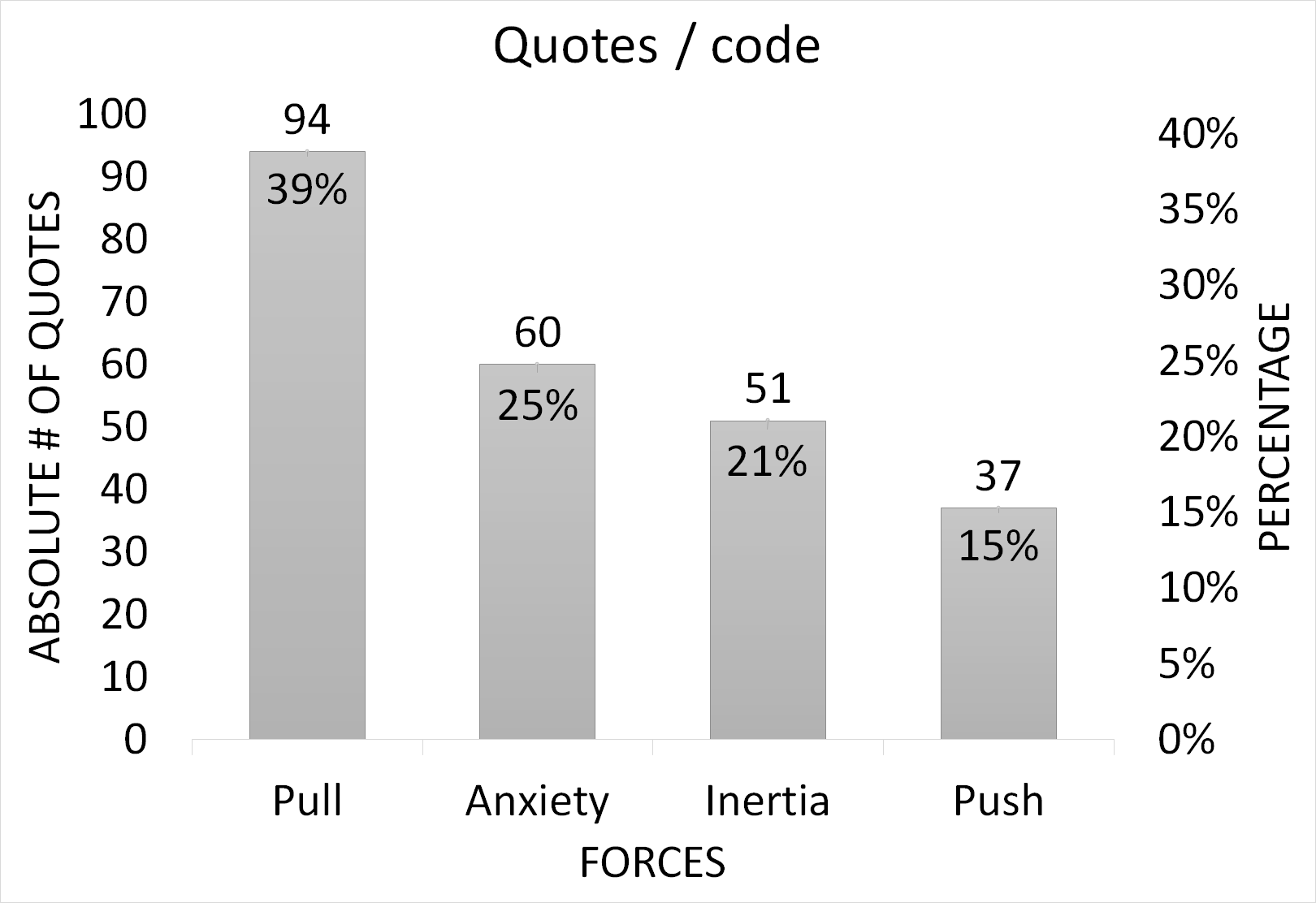}
\caption{Number of quotations related to MBSE adoption forces}
\label{fig:Code_chart}
\end{figure}

\begin{figure}
\centering
\includegraphics[width=\columnwidth]{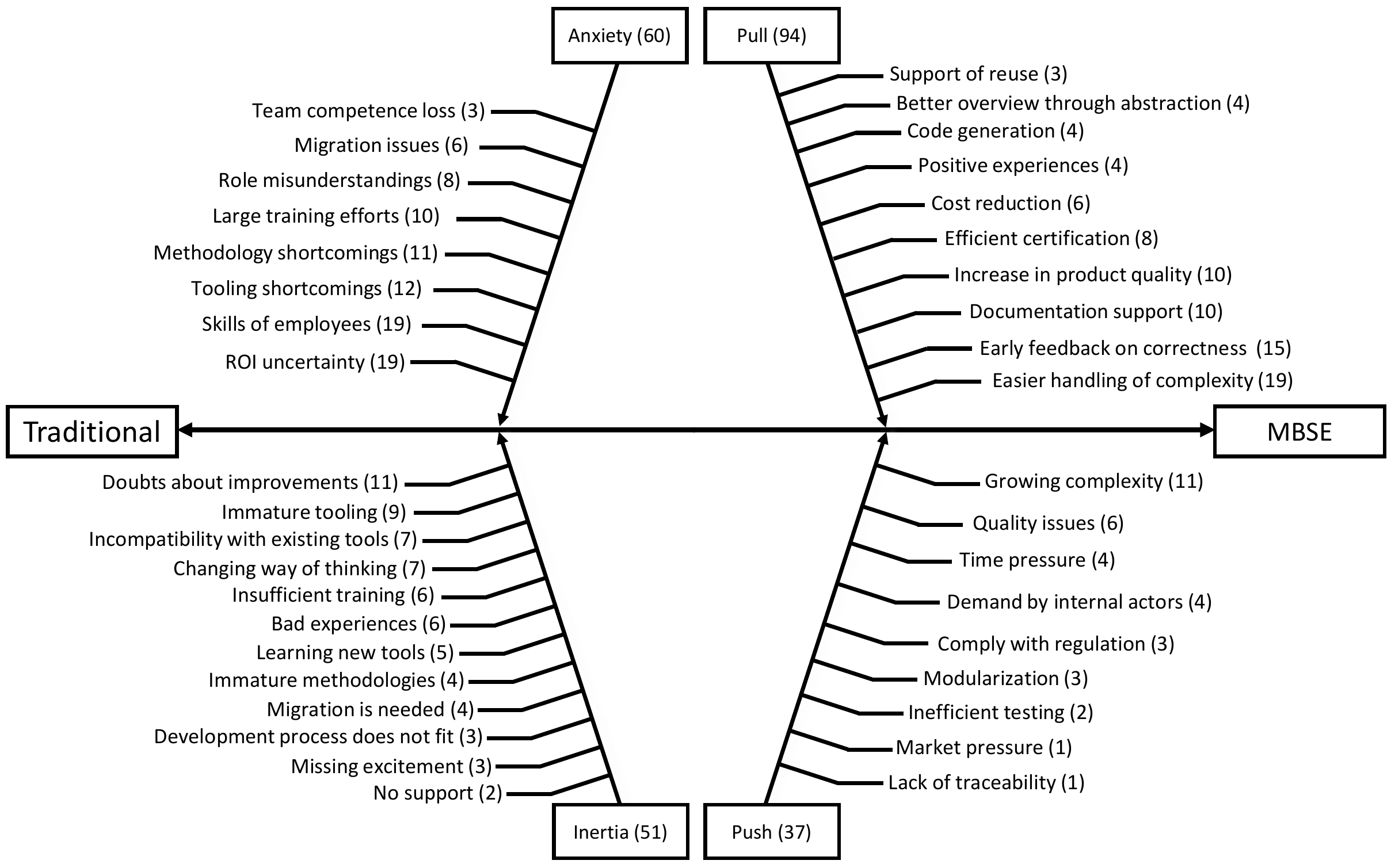}
\caption{Overview of MBSE adoption forces}
\label{fig:Code_overview}
\end{figure}

In total, we coded 242 quotations. Their distribution between the forces can be seen in \figurename~\ref{fig:Code_chart}. The fostering (131 times) and hindering (111) forces were mentioned to a similar amount. Quotations categorized as pull (94) are almost triple of push (37). Comparing both (pull and push), 72\% of the fostering quotations were driven by the benefits of MBSE, while problems in their in-house processes represented 28\%. This can be compared to the number of quotations on inertia (51). Pull forces were coded most, representing 39\% of all quotations. 
To analyze the general attitude of a participant towards MBSE adoption, we divided the number of coded quotations related to fostering forces (push and pull) by the total number of quotations coded for that participant. We considered a participant to have a positive attitude when the ratio of fostering forces was higher than 60\%, a neutral attitude for ratios between 60\% and 40\%, and a negative attitude for a ratio smaller than 40\%. This can be seen in Table~\ref{tbl:participants}. 
The results of the last step of the coding process generated similar codes in different categories (e.g., \emph{Tooling Shortcomings} from Anxiety category and \emph{Immature tooling} or \emph{Incompatibility with existing tools}, both from Inertia category). Although similar names, these codes encompasses disjoint characteristics and their coexistence serves a purpose. All codes created during the analysis can be seen in \figurename~\ref{fig:Code_overview}.

In the following, the preventing forces found in the study are subsequently described and explained using the information from the interview transcripts and the interpretations from coding and analysis.

\subsection{Hindering Force: Inertia}
With 51 distinct quotations, \emph{inertia} forces were mentioned fewer times compared with forces related to anxiety (60 quotations). We structured the inertia related quotations with respect to four inertia topics.

\noindent \textbf{Tooling Inertia (21 coded quotations from 15 interviews)} With 21 quotations, tooling inertia was the most frequently mentioned inertia force. Tooling inertia describes phenomena of the current in-house tooling environment that made our participants refrain from adopting MBSE. Tooling inertia includes resistance against \emph{learning new tools} as well as potential \emph{incompatibilities of MBSE tools} with current tools.
\emph{
``People preferred using Excel instead of the new MBSE tool'' (P8),
``Especially elderly employees who are used to textual specifications have difficulties with drawing tools'' (P15),
``It's not possible to connect\slash trace the models with artifacts in other tools.'' (P5)
}

Apart from the resistance of learning and integrating new tools, our participants reported on resistance of employees if \emph{MBSE tools are immature}. Especially tools with bad user experience, low  stability, and missing basic features are a major factor why employees resist MBSE adoption.
\emph{``Tool is not user friendly. Things are distributed over several menus; you have to look for everything.'' (P5),
``We are working in teams. That's why we need a tool with fine-grained access rights and control.'' (P10)
}

We classified \emph{immature tooling} as inertia force because the expectation that tools are missing important features makes the current situation look not so bad. Tooling issues were also mentioned in the context of anxiety. In that cases, interviewees feared that the available tools cannot fulfill the promises of MBSE.

\noindent \textbf{Context Inertia (18 quotations, 13 interviews)}
A second inertia force mentioned quite often was context inertia, which describes people refraining from MBSE adoption because they believe it does not fit their current business situation. The most mentioned  in this category was doubts about whether MBSE would really improve the current situation. 
\emph{ ``It needs a huge emergency to justify the costs of introducing an MBSE tool.'' (P7),
``Currently, problems are not so urgent yet. Therefore, there is not much willingness to act.'' (P20)}
Another aspect of the context that make people refrain from MBSE adoption is the potential need to migrate old data or legacy systems or when it seems that the current development process does not fit MBSE techniques.
\emph{``Legacy problems are a huge hurdle because, in general, the old way of working must further be maintained and supported.'' (P20),
``MBSE adoption would have caused changes in our development process. Therefore, we didn't do it.'' (P2)}

\noindent \textbf{Personal Inertia (16 quotations, 9 interviews)}
Personal inertia captures forces related to an individual's personality and experiences that hold him\slash her back from adopting MBSE. In our study, these forces were led by the resistance against learning a new way of thinking.
\emph{
``MBSE is not just about changing the notation; it's about changing the way of how I think about systems'' (P2),
``Abstractions in MBSE are not easy to comprehend.'' (P12)}
Similarly, if people had bad experiences with MBSE or related techniques, they have a personal reluctance against adopting MBSE in their current situation.

\noindent \textbf{Maturity Inertia (12 quotations, 8 interviews)}
Maturity inertia was least mentioned in our interviews. Participants were critical about a potential MBSE adoption if they had the impression that the MBSE methodology is not mature enough, there has not been sufficient training before, and there is no support by experts.
\emph{``We first need a common terminology between employees of different departments'' (P7),
``The support for debugging problems is very limited'' (P9)}

\subsection{Hindering Force: Anxiety}
\emph{Anxiety} is a force related to expectations and fears that make MBSE adoption less appealing. These expectations originate from uncertainties that are still to be clarified or a false perception of reality. We structured the anxiety related quotations into the following topics:

\noindent \textbf{ROI Uncertainty (19 quotations, 12 interviews)} Return on investment (ROI) is the benefit resulting from an investment. Introducing MBSE will incur cost spread in several factors such as training, tooling, migration, or lower productivity. Many interviewees were concerned that the investments on introducing MBSE will not pay off. 
\emph{``[It will costs us] A large sum in the million range" (P7), 
``Coaching on the job is very important, but it costs a lot" (P2)}

\noindent \textbf{Skills of Employees (19 quotations, 11 interviews)} Some interviewees fear that (some of) the employees in their company may lack the necessary skills to efficiently adopt MBSE. This can negatively influence the introduction of MBSE in two different ways: Either those employees do not adopt MBSE or they apply them incorrectly.
\emph{``Mechanical engineers know CAD modeling but don't know modeling of behavior" (P1), 
``Modeling should not be an end in itself" (P16)}

\noindent \textbf{Tooling Shortcomings (12 quotations, 8 interviews)} The interviewees perceived problems with tooling as a reason for not introducing MBSE. The interviewees fear that current tool solutions do not address a significant part of the development process and the envisioned benefits of MBSE. Thus, extra work would be necessary to fill the gaps (e.g.,  migration of data between MBSE tools and current tools).
\emph{``Everything in one tool? Nobody wants that" (P5), 
``Performance of the tools [is a challenge for introducing MBSE]" (P7)}

\noindent \textbf{Methodology Shortcomings (11 quotations, 6 interviews)} Many interviewees emphasized the lack of maturity on the current MBSE methodology. This category can be interpreted in two ways. Either the methodology really is incomplete or the knowledge of practitioners is immature. In addition, concerns about the lack of tailored approaches for MBSE introduction were pointed out.  
\emph{``A consistent methodology is lacking, resulting in uncertainties" (P1), 
``There are no process models that integrate MBSE properly." (P11)}

\noindent \textbf{Large Training Efforts (10 quotations, 5 interviews)} This category groups perceived potential problems related to training the team on using MBSE and its respective tools. Some of the codes were related to the costs of training and had intersections with \emph{ROI uncertainty}. Other codes were related to the fear of unsuccessful training. \emph{``Training is necessary: How do I bring my employees to the same level as the experts?" (P7), ``Employees will not accept MBSE if no training is provided before." (P7)}

Besides these major categories, interviewees also mentioned potential \emph{team competence loss} (3 quotations, 3 interviews) and new responsibilities in the team that could cause \emph{role misunderstandings} (8 quotations, 5 interviews). The interviewees perceived \emph{migration issues} (6 quotations, 6 interviews) of projects that started with traditional development method to MBSE.

\subsection{Fostering Force: Push}
With 37 distinct quotations, \emph{push} was the force with the smallest number of quotations. We structured push forces within three categories:

\noindent \textbf{Product Push (20 quotations, 10 interviews)}
We grouped here codes related to product-oriented push forces. \emph{Growing complexity} (11 quotations / 8 interviews) of the software was the code with most quotations within the push forces. As systems become more software-intensive, tackling the growing complexity is currently a real challenge, thus, organizations feel the need to shift to better solutions.
\emph{``Increasing complexity of products [pushes us towards MBSE]" (P1), 
``Complex software, especially with concurrency [pushes us towards MBSE]" (P3)}
Further codes were \emph{quality issues} (6/3) within the product or its specification and the need for \emph{modularization} (3/3) in order to make certification and reuse more efficient.

\noindent \textbf{Stakeholder Enforcement (8 quotations, 4 interviews)}
Some interviewees mentioned that they are forced or at least pushed towards MBSE by recommendations or requests from stakeholders.
\emph{Demands by internal actors} (4/3) such as developers or management push companies towards MBSE adoption as well as legal requirements to \emph{comply with regulations} (3/1). \emph{Market pressure} (1) was mentioned with respect to issues with acquiring talented employees: 
\emph{``We have to be modern, otherwise we will not get good people anymore" (P2)} \looseness=-1

\noindent \textbf{Process Push (7 quotations, 4 interviews)}
Deficiencies of the current process were only mentioned a few times as forces that push companies towards MBSE. The codes were \emph{time pressure} (4/3), 
\emph{inefficient testing} (2/2), and \emph{lack of traceability} (1).
\emph{
``We have no idea what happens when something changes" (P5), ``[We have] Large amounts of requirements; how can the tester handle this?" (P5)}\looseness=-1

In summary, interviewees provided more push forces related to issues with the product instead of issues with the process. 

\subsection{Fostering Force: Pull}
We identified several factors of envisioned benefits that drive companies towards MBSE adoption. 
A majority of the responses given by the interviewees is related to envisioned improvements of the development process. This is interesting since process issues were only mentioned a few times as push factors. 

\noindent \textbf{Easier Handling of Complexity (19 quotations, 12 interviews)} With each new function to integrate, the complexity of software increases. Managing the different software components gets more and more complicated. The interviewees see great opportunities in MBSE to support this challenge. Due to a large number of possible variants of products, complexity of software increases in many companies.
\emph{``[MBSE will help us to] understand highly complex issues or illustrate something" (P15),
``[MBSE will support the] management of product line and variability" (P1)}

\noindent \textbf{Early Feedback on Correctness (15 quotations, 10 interviews)}
The desire for early feedback and front-loading was also a strong pull factor. Especially early verification on higher levels of development were mentioned to improve the development process and finally also the product.
\emph{``Early verification and simulation saves time in the end" (P7),
``[MBSE will provide] better quality due to early fault detection" (P4),
``[MBSE will] Enable automatic verification" (P6)}

\noindent \textbf{Documentation Support (10 quotations, 7 interviews)}
The interviewees expect support to create and manage documentation. The increasing complexity of software development has complicated the management of requirement documents.
\emph{
``[MBSE will provide] better documentation" (P13),
``[MBSE will] generate documentation and code" (P12)}

\noindent \textbf{Increase in Product Quality (10 quotations, 5 interviews)} The interviewees expect better products by introducing MBSE. This includes the final product as well as intermediate development artifacts.
\emph{
``[MBSE will] improve the quality of requirement documents" (P10)}

\noindent \textbf{Efficient Certification (8 quotations, 5 interviews)} Some interviewees envision that MBSE will make it easier to certify software-intensive products. Some interviewees specifically mentioned that MBSE would enable a modular certification, where only parts of the product are certified and not the entire product. 
\emph{
``[MBSE is] necessary to comply with regulatory requirements" (P6),
``[MBSE will enable] modular certification and parallel development" (P6)}

Additional, less frequently mentioned, pull factors include \emph{cost reduction} (6 coded quotations), \emph{positive experiences} (4),  \emph{code generation} (4), \emph{better overview through abstraction} (4), and \emph{support of reuse} (3).


\section{Discussion} \label{label_discussion}
The results show that people from industry have high hopes and expectations for MBSE. However, there are also several hurdles that need to be addressed when adopting MBSE, some of which are very generic. These problems are sometimes even part of the human nature and its natural resistance to change in general. 

\noindent\textbf{Relation to Existing Evidence.}
When comparing our results to related studies on forces of adopting development methodologies in industry, we can identify some general patterns. Hohl~et~al.~\cite{HohlMSS16} report on forces that prevent the adoption of agile development in the automotive domain. They also report on forces of inertia and anxiety resisting a necessary change of mind-set, or limited acceptance for organizational restructuring. Additionally, the current development process was perceived as good-enough. The same forces also appeared in our study. Riungu-Kalliosaari~et~al.~\cite{Riungu-Kalliosaari16} performed a case study on the adoption of DevOps in industry, where they identified five high-level adoption challenges. Three of these challenges were also mentioned as inertia or anxiety factors in our study, namely \emph{deep-seated company culture}, \emph{industry constraints and feasibility}, and \emph{unclear methodology}.
Parallels can also be found in the work of Bauer and Vetr{\`o}~\cite{Bauer16} with respect to the adoption of structured reuse approaches in industry.

Similarly, we also found common and generic goals (i.e., pull forces) that are in the focus of many process improvement activities. Schmitt and Diebold~\cite{Schmitt16} have analyzed common improvement goals that are usually considered when improving the development process. The pull factors that we extracted in our study are part of the main goals elicited by them (e.g., quality and time-to-market). \looseness=-1

When focusing on the forces specific to MBSE that did not appear (so strongly) in the related studies, some factors remain. Incompatibility of MBSE tools with existing tools is a specific inertia force that prevent MBSE adoption. A second force of inertia that was specifically reported for MBSE adoption is the need to adopt a new way of thinking, especially with respect to abstractions. The anxiety forces that we identified were rather generic such that we did not identify any MBSE specific anxiety forces. Interestingly, loss of competences or loss of power, which is a typical anxiety factor, was not mentioned very often.

\noindent\textbf{Impact for Industry.}
MBSE streamlines the activities in all phases of the software lifecycle. It replaces document-based systems engineering and automates several tasks (e.g., code generation). An organization doing the transition from document-based to model-based will require changes in all software development stages, including tools, processes, artifacts, and developing paradigms. 

Our interviewees focused more on push forces related to the product and not so much on the process. One might infer that engineers recognize the growing complexity of their products but they cannot link it to the shortcomings of the current processes. Perhaps, inside their mind, the processes are OK since it has been functioning properly until now and the problem is the product that is getting more difficult to develop. 

The results support decision-making and are an initial step towards efficiently introducing MBSE in companies. Implementing change is always a hassle, therefore companies should manage expectations by setting concrete improvement goals, relating them to concrete MBSE techniques, and making changes step-by-step. Many interviewees mentioned that MBSE adoption should best be piloted in small projects with a clear scope.
%

\noindent\textbf{Impact for Academia.}
MBSE complexity raises uncertainties towards effort and success of its introduction. These uncertainties can be mitigated by knowledge building. Misunderstandings of MBSE, its tools, and processes were quoted many times, which means research is not properly reaching practitioners. This problem is not limited to the MBSE domain but to research in general.
With a clear idea of the forces fostering and hindering MBSE introduction, the next step is to understand how to manage those factors, mitigating them when necessary, or strengthen the ones that contribute to successful MBSE introduction. 
The results provide promising research directions based on real industry needs.

\subsection{Threats to Validity}
The validity of our results is subject to the following threats:

\noindent\textbf{Subject selection bias.} Since this is an exploratory study, we selected a convenience sample of
project partners and personal contacts as study subjects. Although we selected participants from a broad spectrum of companies and industrial domains, the results may be influenced by the fact that all study participants work in Germany. Additionally, the interviewees were selected from an environment where MBSE adoption is a realistic option. 

\noindent\textbf{Researcher bias.} Our study was carried out in the context of a project on transferring MBSE into practice, which means that the authors have a positive attitude towards MBSE in general. Additionally, some of the interviewees are also partners in this project, however, we also interviewed people from companies not involved in the project. To reduce researcher bias, the interviews were conducted by two researchers who took notes independently.

\noindent\textbf{Research method.} 
Validity is threatened by the possibility of misunderstandings between interviewees and the researchers. To minimize this risk, the study goal was explained to the participants prior to the interview. Steps taken to improve the reliability of the interview guide included a review and a pilot test. 
We followed several strategies proposed by Maxwell~\cite{Maxwell12} to mitigate threats. The interviews were conducted as part of a larger project, where we established a \emph{long-term involvement} of the study subjects. As part of this, we presented our study in the context of the project, where the results were reviewed by the project partners.
We substantiate our assertions by providing \emph{quasi-statistics} on the frequency of codes occurrences in the interview data. To validate our results, we \emph{compared} them with existing studies on development methodology adoption. \looseness=-1

\noindent\textbf{External validity.} We expect that our results are representative for the German embedded systems industry, however, we cannot generalize the results to other countries or other types of systems engineering.

\vspace{-0.2cm}
\section{Conclusions}

Organizational change is never easy, especially when trying to introduce complex approaches such as MBSE. In this research, we look for the reasons and factors that prevent or impede companies from adopting MBSE. For this means, we created a forces framework that we used to analyze the information from the verbatim of 20 interviews. We identified forces that hinder and foster MBSE adoption in organizations. We coded the interviews within several discussion rounds. Based on our results, practitioners may challenge their decision processes and adoption strategies. Researchers may study our results and find evidence to quantify and detail the considerations of practitioners. We conclude that bad experiences and frustration about MBSE adoption originate from false or too high expectations. Nevertheless, companies should not underestimate the necessary efforts for convincing employees and addressing their anxiety.

As future work, we plan to analyze the data to investigate correlations between roles and identified categories as well as dependencies between the forces.
Additionally, the research community may create mechanisms to identify the forces within the organization in a more effective and systematic way, analyze how hindering forces can be mitigated, understand how to harvest forces synergy, and figure out which tools and techniques have the highest ROI.

%
%
%
\vspace{-0.2cm}
\section*{Acknowledgements}
This work was partly funded by the German Federal Ministry of Education and Research (BMBF), grant ``SPEDiT, 01IS15058''.

\bibliographystyle{splncs03}
\bibliography{references}
\end{document}